\pdfoutput=1
\documentclass{JINST}

\usepackage{graphicx}

\newcommand{\krm}{$^\mathrm{83m}$Kr}
\newcommand{\kr}{$^\mathrm{83}$Kr}
\newcommand{\kreightyfive}{$^\mathrm{85}$Kr}
\newcommand{\rb}{$^\mathrm{83}$Rb}
\newcommand{\rbkrm}{$^\mathrm{83}$Rb/$^\mathrm{83m}$Kr}
\newcommand{\radona}{$^\mathrm{219}$Rn}
\newcommand{\radonb}{$^\mathrm{220}$Rn}
\newcommand{\radonc}{$^\mathrm{222}$Rn}

\title{A novel $^\mathrm{83m}$Kr tracer method for characterizing xenon gas and cryogenic distillation systems}

\author{S Rosendahl$^a$\thanks{Corresponding
author.}~, K Bokeloh$^a$, E Brown$^a$, I Cristescu$^b$, A Fieguth$^a$, C Huhmann$^a$, O Lebeda$^c$, C Levy$^a$, M Murra$^a$, S Schneider$^a$, D V\'{e}nos$^c$  and C Weinheimer$^a$\\
\llap{$^a$}Institut f\"ur Kernphysik, University of M\"unster,\\
  Wilhelm-Klemm Stra\ss e 9, 48149 M\"unster, Germany\\
\llap{$^b$}Tritium Laboratory Karlsruhe, Karlsruhe Institute of Technology,\\
  Hermann Von Helmholtz-Platz 1, 76344 Eggenstein-Leopoldshafen, Germany\\
 \llap{$^c$}Nuclear Physics Institute, Academy of Sciences of the Czech Republic,\\
  CZ 250 68, \v{R}e\v{z} near Prague, Czech Republic\\ 
  
  E-mail: \email{rosendahl@wwu.de}}

\abstract{The radioactive isomer \krm\, has many properties that make it very useful for various applications. Its low energy decay products, like conversion, shake-off and Auger electrons as well as X- and $\gamma$-rays are used for calibration purposes in neutrino mass experiments and direct dark matter detection experiments. Thanks to the short half-life of 1.83\thinspace h and the decay to the ground state $^{83}$Kr, one does not risk contamination of any low-background experiment with long-lived radionuclides. In this paper, we present two new applications of \krm . It can be used as a radioactive tracer in noble gases to characterize the particle flow inside of gas routing systems. A method of doping \krm\, into xenon gas and its detection, using special custom-made detectors, based on a photomultiplier tube, is described. This technique has been used to determine the circulation speed of gas particles inside of a gas purification system for xenon. Furthermore, \krm\,   can 
be used to rapidly estimate separation performance of a distillation system. 
}


\begin{document}

\section{Introduction}

The number of applications of the isomer \krm , produced by the decay of \rb\ (T$_{1/2} = 86.2$\thinspace d) via electron capture (see \cite{nuclear_chart_bnl}), in nuclear, particle and astroparticle physics has increased in recent years. Since it produces mono-energetic conversion electrons 
from the highly converted gamma transitions of $E_{\gamma,1} = 32.2$\thinspace keV and $E_{\gamma,2} = 9.4$\thinspace keV, it has been used as a calibration source in neutrino mass experiments at Los Alamos, Mainz, Troitsk \cite{robertson,picard92b,bornschein,aseev} and in KATRIN 
\cite{KDR,zboril, mos}. 
The isomer \krm\ has been applied for the calibration of the time projection chamber (TPC) of the ALICE detector at CERN, which is the world largest TPC with a volume of about 90~m$^{3}$ \cite{alice}.
The first studies also demonstrated excellent suitability of $^{83\mathrm{m}}$Kr for liquid noble gas detectors \cite{baudis,plante}. 

\krm\ does not contaminate low-background detectors, since it has short half-life (T$_{1/2} = 1.83$\thinspace h)  and decays
into the stable ground state \kr .
It has been, therefore, introduced as an internal calibration source in direct dark matter experiments using liquid noble gases (e.g. LUX \cite{lux}, DarkSide-50 \cite{darkside50}) and it is planned as a calibration source for XENON1T \cite{xenon1t}. 

Another \krm\ application as a tracer in xenon gas is presented in this paper: Emanating from a \rb\ generator, \krm\ is mixed into xenon gas. The  \krm -decay in the xenon gas is detected with a simple detector system measuring the xenon scintillation light with a photomultiplier tube (PMT). 

For the next generation of direct dark matter detection experiments using liquid noble gas detectors, it is essential to reduce radioactive background substantially. One of its components originates in intrinsic contamination of the noble gas itself.
For xenon detectors, it is predominantly the isotope \kreightyfive\ produced in above-ground nuclear bombs testing and in nuclear reactors, as well as the radon isotopes \radona , \radonb\ and \radonc\ originating from the actinium, thorium and uranium decay chains, which are present in trace amounts in the detector materials.
Argon detectors can distinguish most of these components by pulse shape discrimination.  In argon detectors, the cosmogenically produced isotope $^{39}$Ar forms the most relevant intrinsic contamination.

The most common technology to remove krypton from xenon is cryogenic distillation \cite{Abe09, Wang14}, where concentrations of $^{\mathrm{nat}}$Kr/Xe $<$ 20\thinspace ppt\footnote[1]{1\, ppt$ = 10^{-12}$\, mol/mol} have been reached \cite{Aprile1}. For further reduction of the intrinsic radioactive background, the krypton level has to be still decreased, i.e. the distillation process has to be improved. Naturally, an appropriate diagnostic tool is needed to characterize new distillation systems. The state-of-the-art gas analysis systems allow to monitor concentrations of $^{\mathrm{nat}}$Kr/Xe $<$ 1\thinspace ppt, using gas chromatography combined with noble gas mass spectrometry \cite{lindemann} or atom trap trace analysis \cite{aprile_atta}. Such analytical methods are not suitable for dynamic studies of distillation systems, since they are not enough rapid due to time required for both sample preparation and analysis itself, what makes continuous operation monitoring very difficult. We, therefore, 
developed a method of 
doping the gas with trace amounts of radioactive $^{83\mathrm{m}}$Kr and measured its decay using special photomultiplier tubes that have quantum efficiency optimized for the scintillation light from gaseous xenon at 171\,nm. Our measuring procedure should also be feasible for other noble gases like argon or neon, however, introducing wavelength shifters or specialized photomultipliers will be necessary.

In the section 2 of this paper, we present the \krm -mixing method and the detector setup to measure the \krm -decay rate in xenon gas. In the section 3, we 
describe the \krm\ mixing into xenon gas and the characterization of a xenon gas system. In section 4, we apply our mixing and detection technique to characterize the efficiency of a cryogenic distillation system for removal of krypton from xenon.

\section{The $^{83\mathrm{m}}$Kr tracer method}

\subsection{$^\mathrm{83m}$Kr decay}
\begin{figure}[!!!h]
\center
\includegraphics[width=0.6\textwidth]{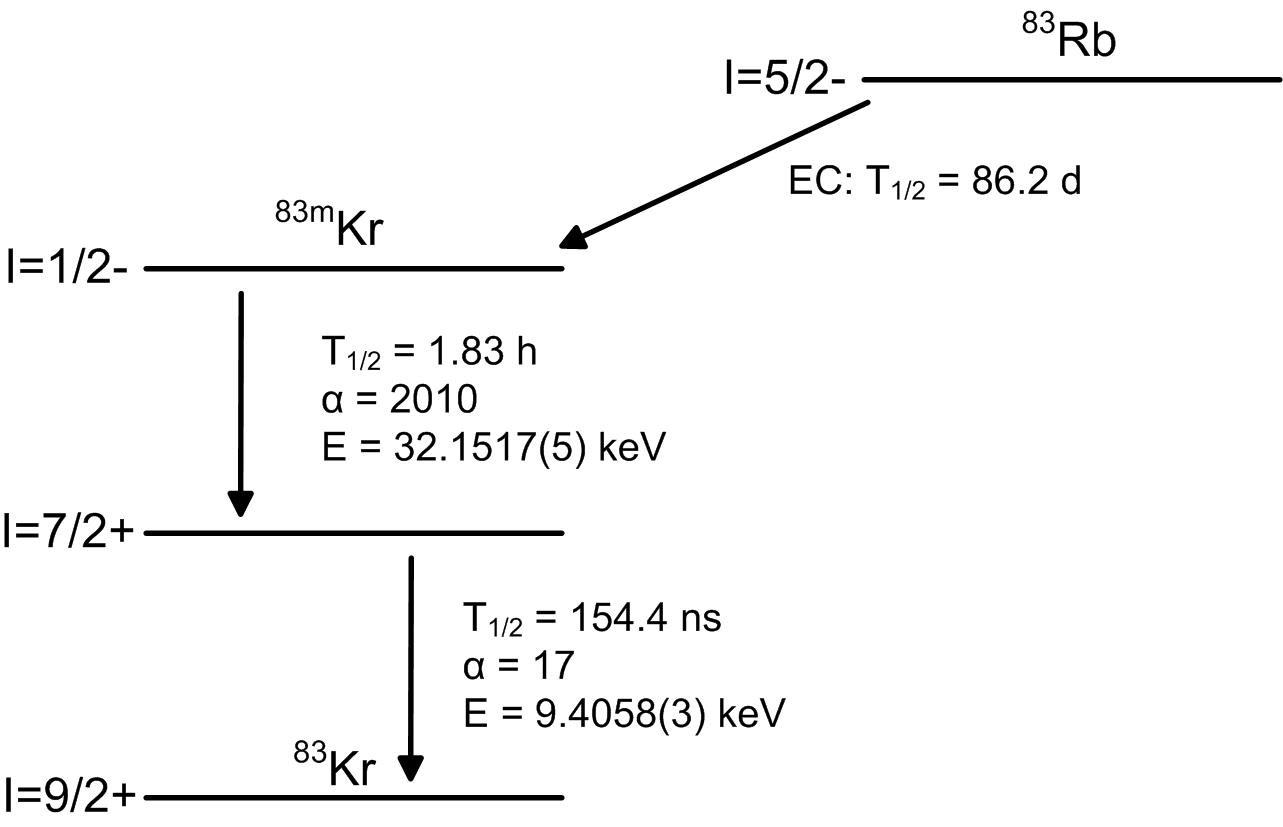}
\caption{\textbf{Decay scheme of $^{83}$Rb to $^{83}$Kr.} The simplified decay scheme of $^{83}$Rb to $^{83\mathrm{m}}$Kr, which further decays via two highly converted transitions (total conversion coefficients $\alpha = 2010$ and $\alpha = 17$ respectively) into the stable $^{83}$Kr. The scheme is based on \cite{nuclear_chart_bnl}, \cite{ven06} and \cite{Slezak12}.}
    \label{fig:rb_decay}
\end{figure}

The isomeric state$^{83\mathrm{m}}$Kr at an excitation energy of 41.534~keV is populated by the electron capture of $^{83}$Rb with a branching ratio of 77.9~\%, from which it decays to the ground state of \kr\ by two highly converted transitions. The intermediate state at 9.4058~keV has a very short half-life of 154.4~ns. Therefore, the half-life of 1.83~h of the isomeric state governs decay kinetics. Figure \ref{fig:rb_decay} displays the decay scheme. 
Significant advantage of using $^{83\mathrm{m}}$Kr is that the half-life of 1.83 hours is short enough to produce a rather high signal rate in the decay detectors, 
as well as still long enough for $^{83\mathrm{m}}$Kr distribution inside of gas and distillation systems. Furthermore, this isomer decays solely via isomeric transition to its stable ground state $^{83}$Kr, so there is no risk of contamination of xenon with long-lived decay products. The only potential risk of radioactive contamination represents the mother nuclide $^{83}$Rb with half-life of 86.2~d, however, this contamination can be efficiently prevented (see below). 

\subsection{Doping of xenon gas with $^{\mathrm{83m}}$Kr}
\begin{figure}[!!!h]
\center
\includegraphics[width=0.55\textwidth]{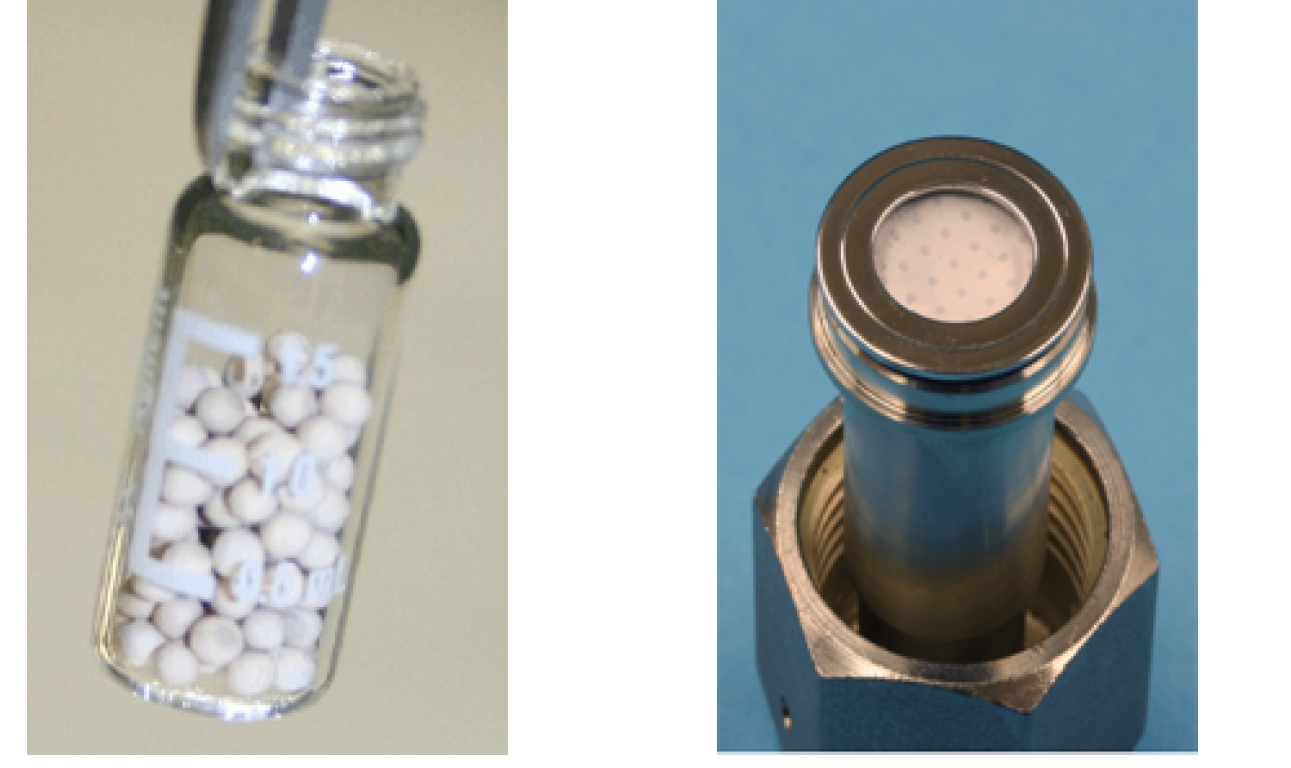}
\caption{\textbf{Zeolite beads and the source holder.} The zeolite beads with absorbed $^{83}$Rb: the source inside of the glass transport vial (left) and in the stainless steel cylinder with a VCR connector for attachment to the xenon system covered by protective PTFE membrane filter (right).}
    \label{fig:pellets_rb_generator}
\end{figure}

The radionuclide \rb\ was produced on the cyclotron U-120M at the Nuclear Physics Institute, Academy of Sciences of the Czech Republic, by irradiating a medium-pressurized (13~bar) natural krypton gas target by a 26~MeV proton beam with an intensity of 15~$\mu$A. After irradiation, the rubidium isotopes were washed out of the
target walls by several portions of deionized water almost quantitatively. The volume of the resulting solution was then reduced by evaporation in a quartz beaker under an infrared lamp. Finally a portion of the \rb\ solution with the desired activity was absorbed in zeolite beads of 2~mm diameter having 0.5~nm pores (see figure \ref{fig:pellets_rb_generator} left). The beads were dried under infrared lamp and then for 2~h at $350^\circ$C \cite{venos_2005}.

The structure and ion-exchange properties of the zeolite allow efficient emanation of \krm\ , while \rb\ is strongly trapped. The possible release of \rb\ from the beads has been investigated in \cite{Hannen1}, but no detectable traces of \rb\ were observed (less than 1 mBq after 2 weeks measurement of 1.8 MBq source). The $^{83\mathrm{m}}$Kr generators based on $^{83}$Rb adsorbed on zeolite can be, therefore, used in low count rate experiments like XENON100 or KATRIN without the contamination risk.

The emanation efficiency of $^{83\mathrm{m}}$Kr from zeolite is function of time and environment around the source. A freshly prepared source emanates about 80\% of $^{\mathrm{83m}}$Kr born in the $^{83}$Rb decay. If exposure to air continues, emanation efficiency drops to ca 15\% within 3--5 days. When such a source is exposed to the vacuum of $10^{-2}$ mbar or better, the $^{\mathrm{83m}}$Kr emanation efficiency increases to ca 70\% and stays constant for more than a week. The restoration effect cannot be achieved in case of rotary pump vacuum. 

The zeolite beads were placed into a stainless steel container linked by a VCR connector to a xenon gas system (see figure \ref{fig:pellets_rb_generator} right). A PTFE membrane filter with a pore size of 220~nm and a grid, closing the stainless steel container, prevent the potential release of zeolite microparticles from entering the system, while allowing easy transport of \krm\ . The \krm\ atoms are mixed with the xenon gas by diffusion. When xenon gas is circulating in the system, the \krm\ atoms follow the circulation as well (see also section \ref{sec:gas_system}).

\subsection{Detection of the $^\mathrm{83m}$Kr in gaseous xenon}

\begin{figure}[!!!h]
\center
\includegraphics[width=0.8\textwidth]{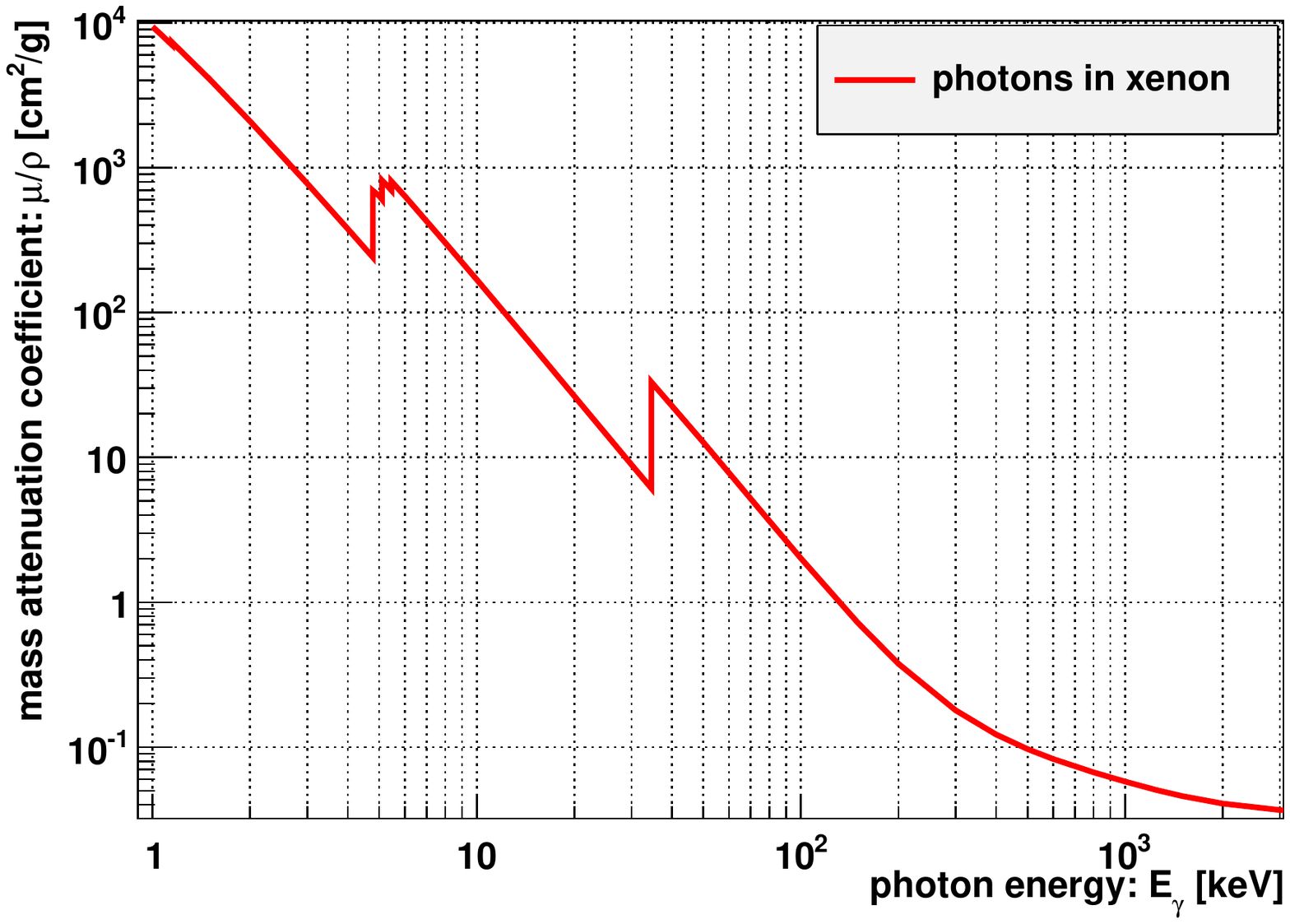} \hfill \includegraphics[width=0.8\textwidth]{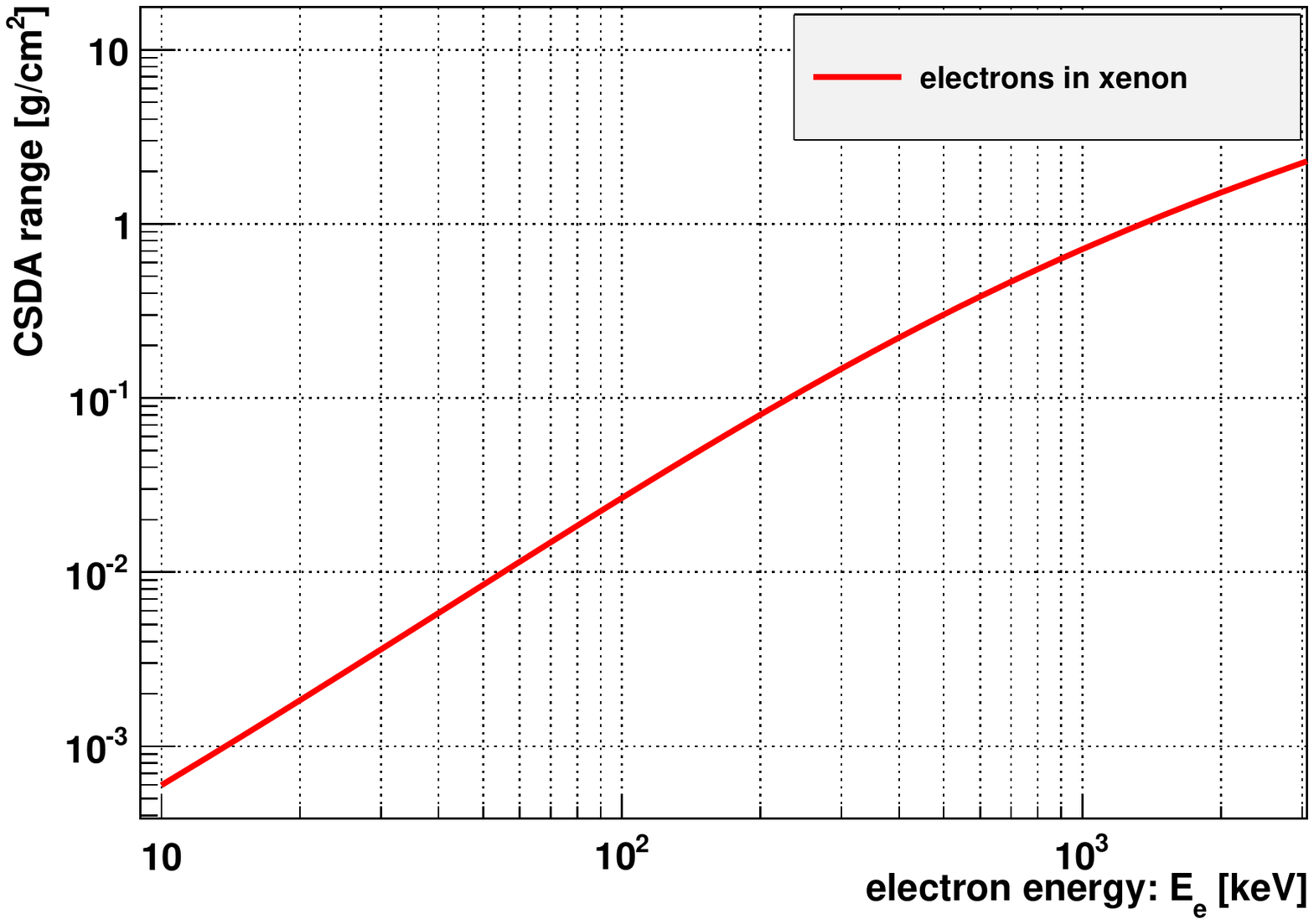}
\vspace*{0.1cm}

\caption{\textbf{Mass attenuation coefficients of photons and CSDA range of electrons in xenon.} Top: Mass attenuation coefficient $\mu/\rho$ for photons with energies up to 3\,MeV, taken from \cite{NIST1}. Bottom: CSDA range of electrons in xenon for energies up to 3\,MeV, data taken from \cite{NIST2}. The xenon gas density at room temperature is 5.3\,mg/cm$^{3}$, while liquid xenon density at T=$-100^\circ$C and p=1.6\,bar is 2.8\,g/cm$^3$.}
    \label{fig:attenuation}
\end{figure}

The radioactive decay of $^{83\mathrm{m}}$Kr to $^{83}$Kr can be detected by the interaction of the decay products (conversion, shake-off and Auger electrons as well as X- and $\gamma$-rays) with gaseous xenon that produces scintillation light of 171\thinspace nm wavelength (VUV light) \cite{Takahashi1}. 
The scintillation light in gaseous xenon is produced by the de-excitation of xenon dimers that are formed by the recombination of an excited xenon atom with another Xe atom in the ground state \cite{Aprile2}. Since these dimers cannot exist in their ground state, xenon itself is transparent for its scintillation light. It can be consequently detected by photomultiplier tubes (PMTs) with a photocathode adapted to this wavelength. The measured rate is 
proportional to $^{83\mathrm{m}}$Kr density in xenon gas. 

Figure  \ref{fig:attenuation} shows the mass attenuation coefficient of X- and $\gamma$-rays and the CSDA range of electrons in gaseous xenon. For typical energies of up to 30~keV, the range of electrons is less than 1\,cm in xenon gas at a pressure of 1~bar and the mean free path of the X- and $\gamma$-rays ranges from 20\,cm for 30\,keV to 1\,cm at 10\,keV. This defines necessary size of a \krm\ decay detector, based on the detection of scintillation light in gaseous xenon.

\subsection{The detector design}

\begin{figure}[!!!h]
\center
\includegraphics[width=0.45\textwidth]{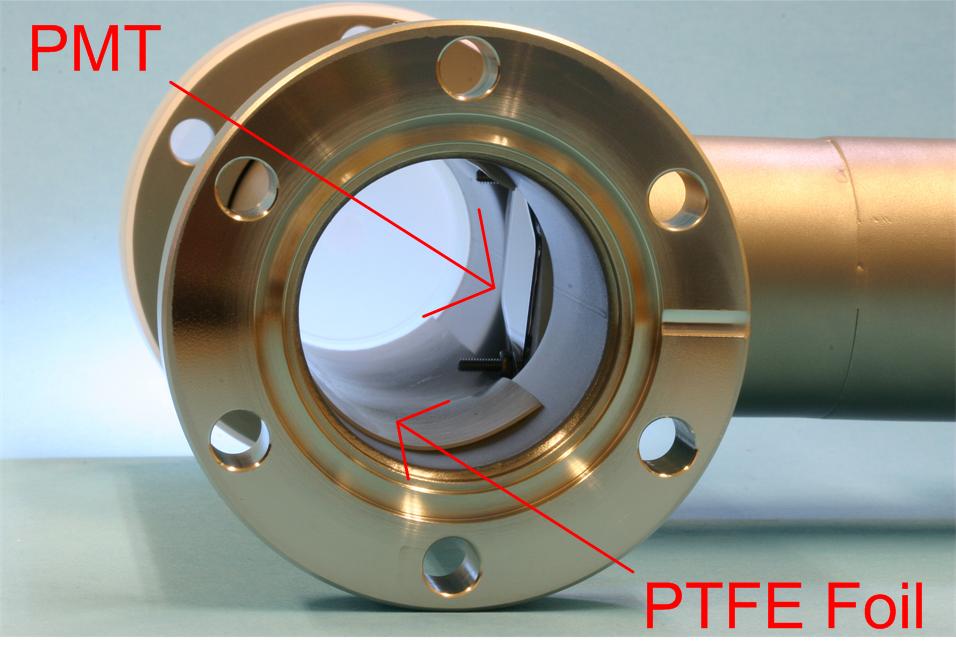} \hfill \includegraphics[width=0.45\textwidth]{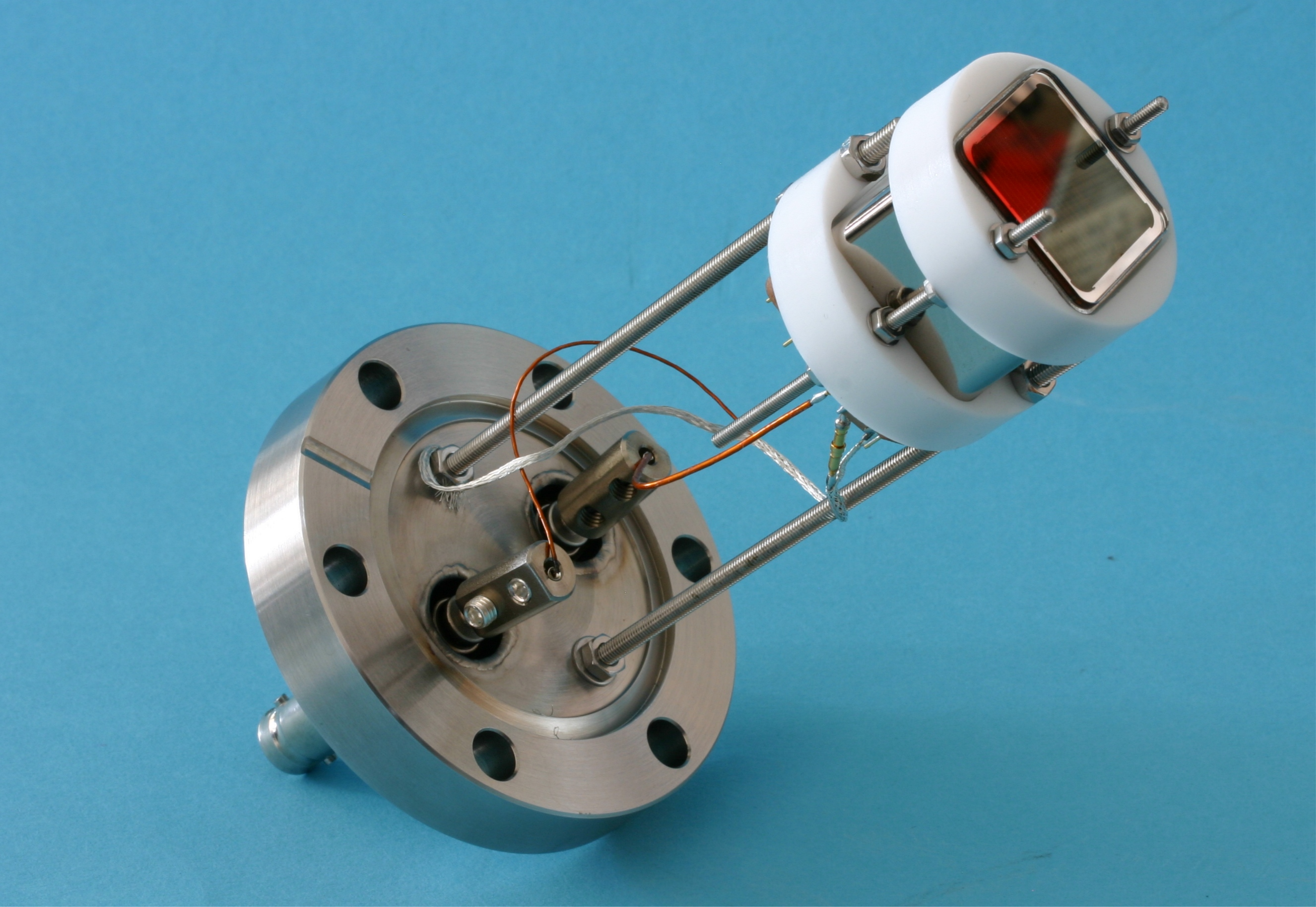}
\vspace*{0.2cm}

\includegraphics[width=0.45\textwidth]{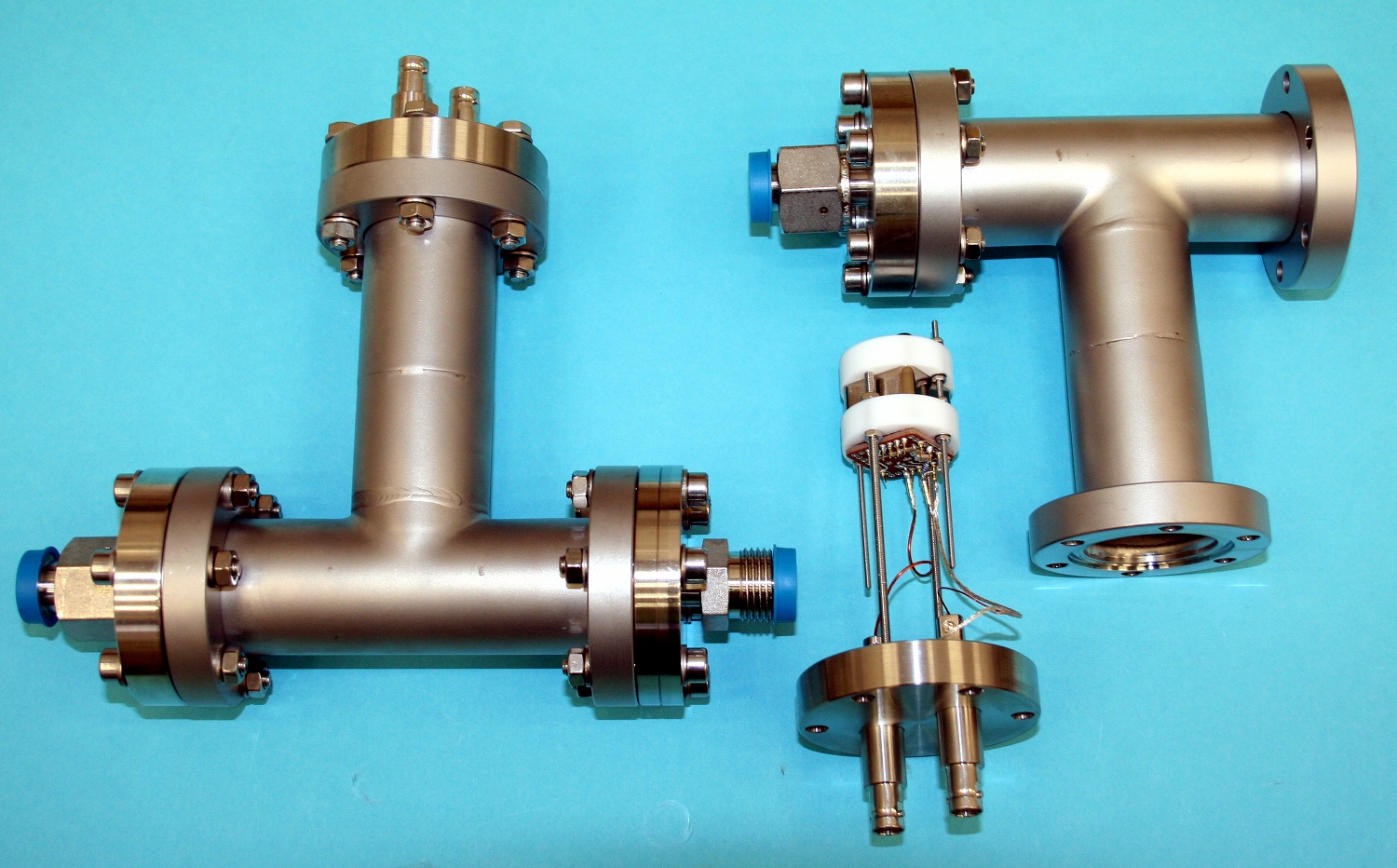}
\caption{\textbf{Custom-made \krm\ decay detector.} Upper left: PMT mounted inside a CF-40 T-piece and the arrangement of the PTFE reflector foil.
Upper right: PMT mounted on the CF-40 flange with feedthroughs. The mounting of the PMT with the teflon holder to the CF-40 flange is displayed. The flange is equipped with high voltage and signal feedthroughs to operate the PMT.
Bottom: \krm\ decay detectors inside of the xenon gas tubings.}
    \label{fig:kr83m_decay_detector}
\end{figure}

To detect the scintillation light formed in xenon, a custom detector has been constructed: A 1~inch photomultiplier tube (PMT, type R8520-06-AL, Hamamatsu, Japan) is mounted perpendicularly to a stainless tube of 40~mm diameter to monitor the xenon that passes through the line (see figure \ref{fig:kr83m_decay_detector} upper left). 
The bialkali photocathode of the PMT provides a quantum efficiency of $\geq 30$~\% at $\lambda=178$\,nm. 
A Polytetrafluoroethylene (PTFE) foil inside the stainless steel tube enhances the reflectivity of the walls in order to increase number of scintillation photons detected by the PMT.
The PMT is surrounded by a holding structure made out of PTFE to protect the sensitive detector and also to hold it in a defined position. It is attached to a CF-40 flange with threaded rods (see figure \ref{fig:kr83m_decay_detector} upper right). 
The flange contains two SHV feedthroughs for a high voltage supply to the PMT (+800\thinspace V) and for a read out of the PMT signals. The casing of the PMT is at ground potential.
The CF-40 flange is then mounted to a CF-40 T-piece on the outgoing line, while xenon can be flushed through the straight end. The detector is connected to the gas system by using half inch VCR connectors (see figure \ref{fig:kr83m_decay_detector} bottom).
 
Our measurements revealed necessity, in order to achieve high light collection efficiency, to clean the xenon gas by a zirconium-based hot getter (type Monotorr, SAES) from impurities that absorb the scintillation light. We continuously run the xenon gas through the getter and gas purity was monitored by measuring xenon humidity with use of a moisture analyzer (type HALO+, Tiger Optics).
In order to minimize the impurities that are introduced to the xenon, most of the parts have been cleaned in an ultrasonic bath using an alkaline degreaser (p3-almeco, Henkel) followed by deionized water. The parts that cannot be cleaned in the ultrasonic bath, like cables or bases, have been rinsed with pure ethanol.

The tubing volume, monitored by the PMT, is of cylindrical shape with length of $\approx$10.5\,cm.
Since the energy is well deposited on short distances, one can estimate the number of scintillation photons produced in xenon due to single \krm\ decay.  The average energy deposition necessary for formation of one electron-ion pair is  $W = 22.0$\,eV for gaseous xenon \cite{Aprile2}.
The average number of photons N$_{\gamma}$ can be estimated as the fraction of deposited energy E$_{0}$ and the average energy per electron-ion pair $W$. For the I=1/2- to I=7/2+ transition with 32.2\,keV energy difference, N$_{\gamma, 32keV}$=E$_{0}$/W$_{\mathrm{xenon}}=1460$ and for the I=7/2+ to I=9/2+ transition with 9.4\,keV,  N$_{\gamma, 9keV} = 427$. In liquid xenon it has been observed that the number of photons on the latter decay is partly quenched due to the ionization following the first decay \cite{plante}. 
From the summed energy ($\sim$41~keV) of the cascade $\approx1900$ photons can be produced, which should be more than enough for a detection of every \krm\ decay with high efficiency also considering the quantum efficiency of the PMT, scintillation light absorption by impurities, loss of light by non-perfect reflection of the walls, etc.
 
\begin{figure}[!!!h]
\center
\includegraphics[width=0.7\textwidth]{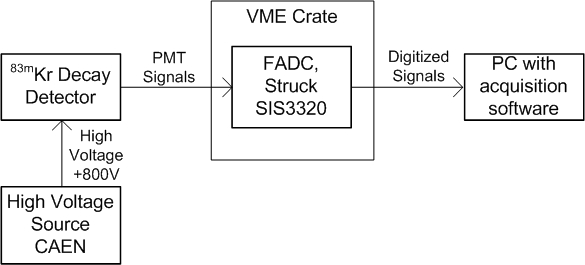}
\caption{\textbf{Schematic of the data acquisition system.}  The PMT is supplied with high voltage, while the raw signals are acquired with a FADC and processed by a PC using a VME interface to the digitizer. }
    \label{fig:daq}
\end{figure}

The PMT signals are acquired using an eight channel flash analog to digital converter (FADC) (type SIS3320, Struck Innovative Systeme) without any fast amplifiers (see figure \ref{fig:daq}). The FADC provides a 12 bit resolution and was used with 100\,MHz sampling rate. It can be readout by a VME interface connected to the lab PC with a custom acquisition program. For the PMT supply voltage of +800\,V a NIM module (type N470, CAEN) was used.

\begin{figure}[!!!h]
\center
\includegraphics[width=0.8\textwidth]{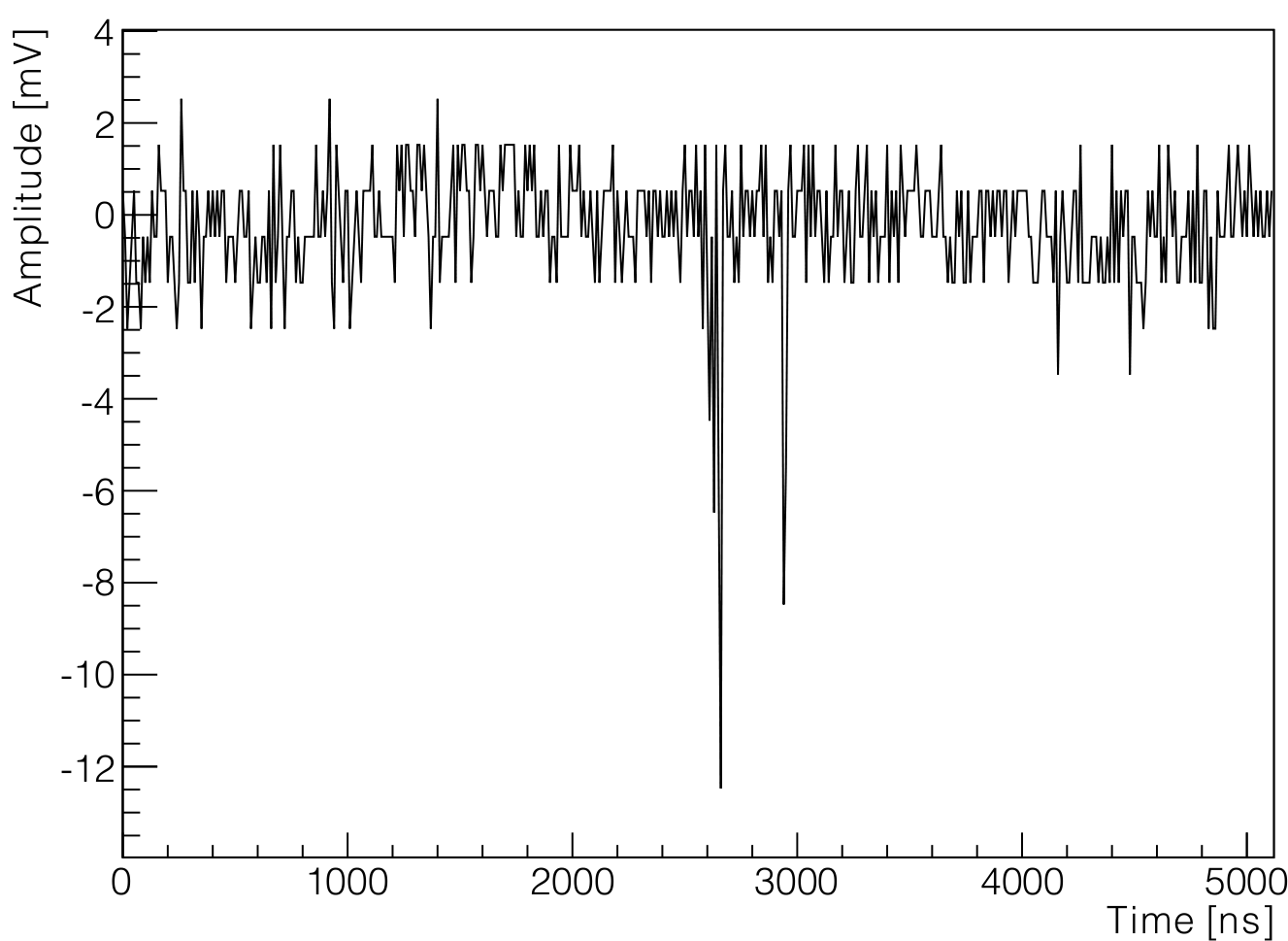}
\caption{\textbf{Example of a double coincidence waveform} This plot shows the waveforms that are acquired from the $^{83\mathrm{m}}$Kr decay detectors. In this waveform, the double coincidence signal represents the two resolved transitions in the $^{83\mathrm{m}}$Kr decay.}
    \label{fig:kr_decay_double_coincidence}
\end{figure}

Since scintillation light pulses in xenon are very fast, it is possible to resolve the two transitions in the \krm\ decay. A typical waveform of a \krm\ decay event is shown in figure \ref{fig:kr_decay_double_coincidence}. The pulse heights of the two transitions are different due to the different energies of the decay steps (see figure \ref{fig:rb_decay}).

\section{Characterizing the mixing of $^\mathrm{83m}$Kr in a xenon gas system}
\label{sec:gas_system}
\begin{figure}[!!!h]
\center
\includegraphics[width=0.7\textwidth]{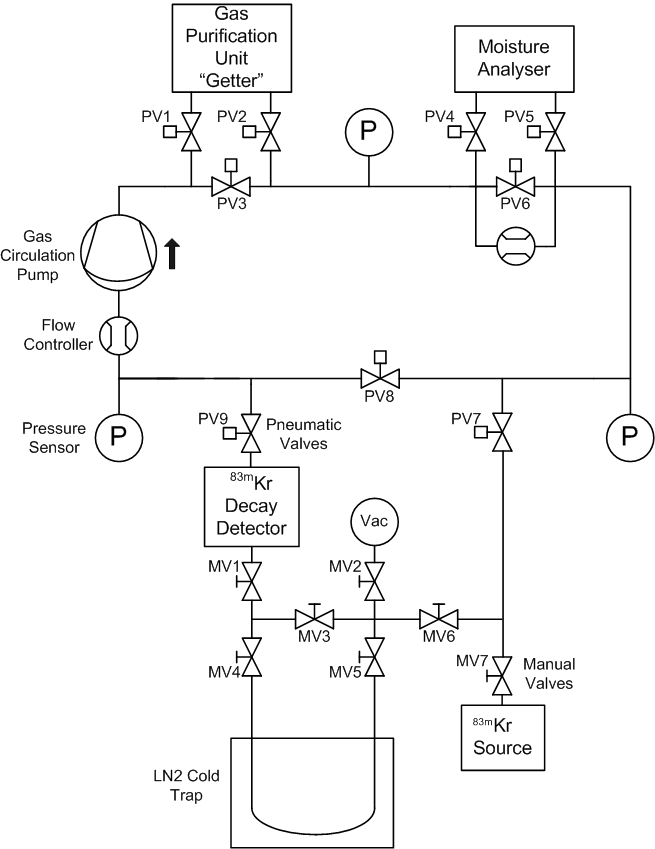}
\caption{\textbf{Flowchart of the gas circulation system.} The simplified scheme of the gas circulation system, set up at IKP M\"unster. For the clear arrangement of the scheme, several additional ports to other subsystems, which have been closed during the measurements, have been neglected. The black arrow indicates the direction of the gas flow.}
    \label{fig:gas_system}
\end{figure}

The \rb\, source, emanating \krm,\, and the \krm\, decay detector were attached to a xenon gas system, which allowed for circulating the gaseous xenon. The scheme of the setup is shown in figure \ref{fig:gas_system}. This system has been set up as a demonstrator for a gas distribution and purification system for the dark matter experiment XENON1T.  

The gas circulation through electro-polished stainless steel lines of half-inch outer diameter is driven by a membrane pump (type N143AN.12E, KNF) and controlled with a mass flow controller (type 1579A, MKS), calibrated for xenon. The pressure inside the system is measured at several locations by using capacitance manometer (type 121A, MKS). The routing of the gas is controlled with bellow-sealed valves, either remotely controlled, pneumatic valves (type SS-8BG-VCR-5C, Swagelok, in figure \ref{fig:gas_system} denoted by PV1 to PV9) or manually operated valves (type SS-8BG-VCR, Swagelok, in figure \ref{fig:gas_system} denoted by MV1 to MV7). The gas can be routed through the getter (PV3 closed, PV1 and PV2 open), while the moisture analyzer is mounted to a bypass after the getter (PV6 closed, PV4 and PV5 open) and allows to measure the fraction of water in xenon down to 200\thinspace ppt. Since the device needs a very low flow rate of $\approx$1\,slpm of xenon, compared to the flow rate of 8\,slpm during 
normal 
operation, the majority of the gas passes the analyzer through the flow controller.

The valves MV3 - MV5 allow the gas routing to the $^{83\mathrm{m}}$Kr decay detector either directly, or via a tube, immersed in a liquid nitrogen bath. The latter one can be used to test the temperature dependence of the krypton collection efficiency. The source also has an additional valve (MV7) to separate it completely from the system. 
This section of the system is also connected to a vacuum pump (Vac) by the valve MV2 which allows for the evacuation of these volumes independently from the rest of the circuit.

\subsection{Performance of the system}

The pneumatic valves PV7 and PV9 were closed, while PV8 remained open and xenon was circulating through the closed circuit and the getter. Before the collection was started, the volume above the $^{83}$Rb containing zeolite beads was evacuated to $\approx 3\cdot10^{-3}$\,mbar, by opening MV2-MV6, while MV1 and MV7 remained close. For the collection, MV3-MV6 remained open. MV2 was closed, while MV7 was opened and $^{83\mathrm{m}}$Kr, emanating from the beads, was collected in the section for 6 hours. 

\begin{figure}[!!!h]
\center
\includegraphics[width=1.0\textwidth]{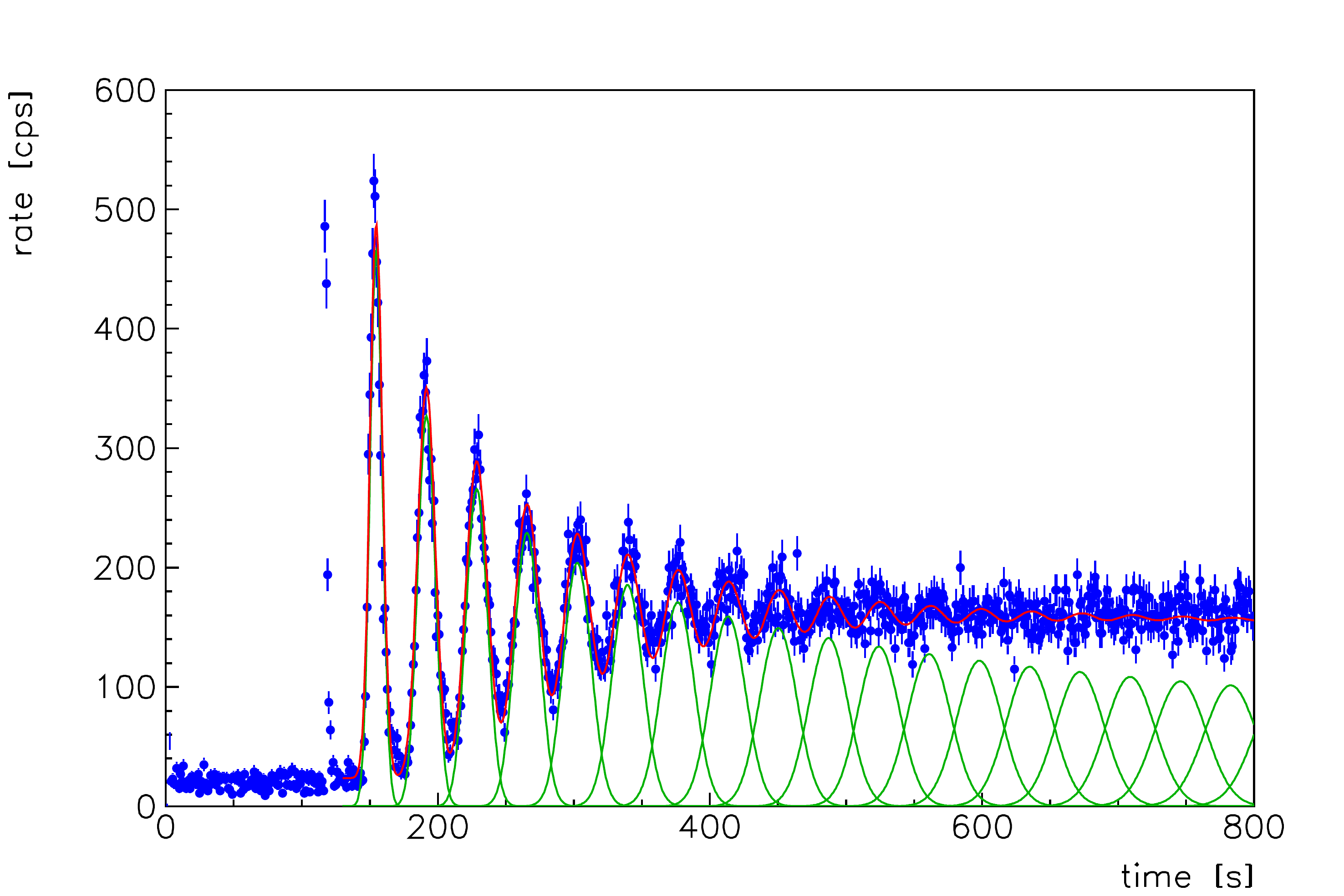}
\caption{\textbf{\krm\ decay rate in closed loop after injecting \krm\ at $T=118$~s.} The fit described in equation 3.1 is shown in red, while the individual single Gaussian terms of equation 3.1 are displayed in green.}
    \label{fig:timeevolution}
\end{figure}

After the collection, the valve to the source MV7 was closed. The pneumatic valves PV7 and PV9 as well as MV1 were opened and PV8 was closed. The $^{83\mathrm{m}}$Kr then mixed with the circulating xenon, which was routed past the detector.
On figure \ref{fig:timeevolution}, the time evolution of a measurement of the first 800\,s is shown. After the opening of the valve the signal rate peaks to a maximum of $\approx 500$\,cps and than decreases in a damped oscillation until it reaches a constant rate of $\approx 160$\,cps on the detector. The oscillation is a function of the xenon circulation flow rate. Since the krypton cloud is transported by the xenon flow around the closed loop, it forms a bolus in the beginning that diffuses with time and asymptotically approaches a uniform $^{83\mathrm{m}}$Kr concentration in xenon circuit, while slowly decaying.

We see the \krm -decays in the detector when the
\krm\ passes by for the first time at $t_0 \approx 118$~s. Its
decay gives rise to a $\delta$-distribution-like peak. After having cycled one time through the setup, the \krm\
passes by the \krm -decay detector at $t_0 + T_{per} \approx 155$~s. Now the time distribution already has a Gaussian shape with a width $\sigma$ due to diffusion of the \krm\ in the xenon. In fact, the process that is described here, is a diffusion process in a turbulent gas stream, driven by the circulation pump. To describe the data, we assume a further Gaussian broadening by $\sigma$ and a further delay of the peak by $T_{per}$ per cycle. The overall rate is multiplied by an exponential function to describe the radioactive decay of \krm, see the definition of the function $f(t)$ in equation \ref{eq:fit}. We fit this function $f(t)$ to the data from the first Gaussian peak on, see figure \ref{fig:timeevolution}.

\begin{eqnarray}
f(t)=a+\sum_{n=1}^N \frac{b}{\sqrt{2\pi n} \cdot \sigma} \cdot \exp\{- \frac{(t-(t_0 + n \cdot T_{per})^2}{2\cdot (\sigma \cdot \sqrt{n})^2}\} \cdot \exp(-t/ \tau)
\label{eq:fit}
\end{eqnarray}

\begin{figure}[!!!h]
\center
\includegraphics[width=0.8\textwidth]{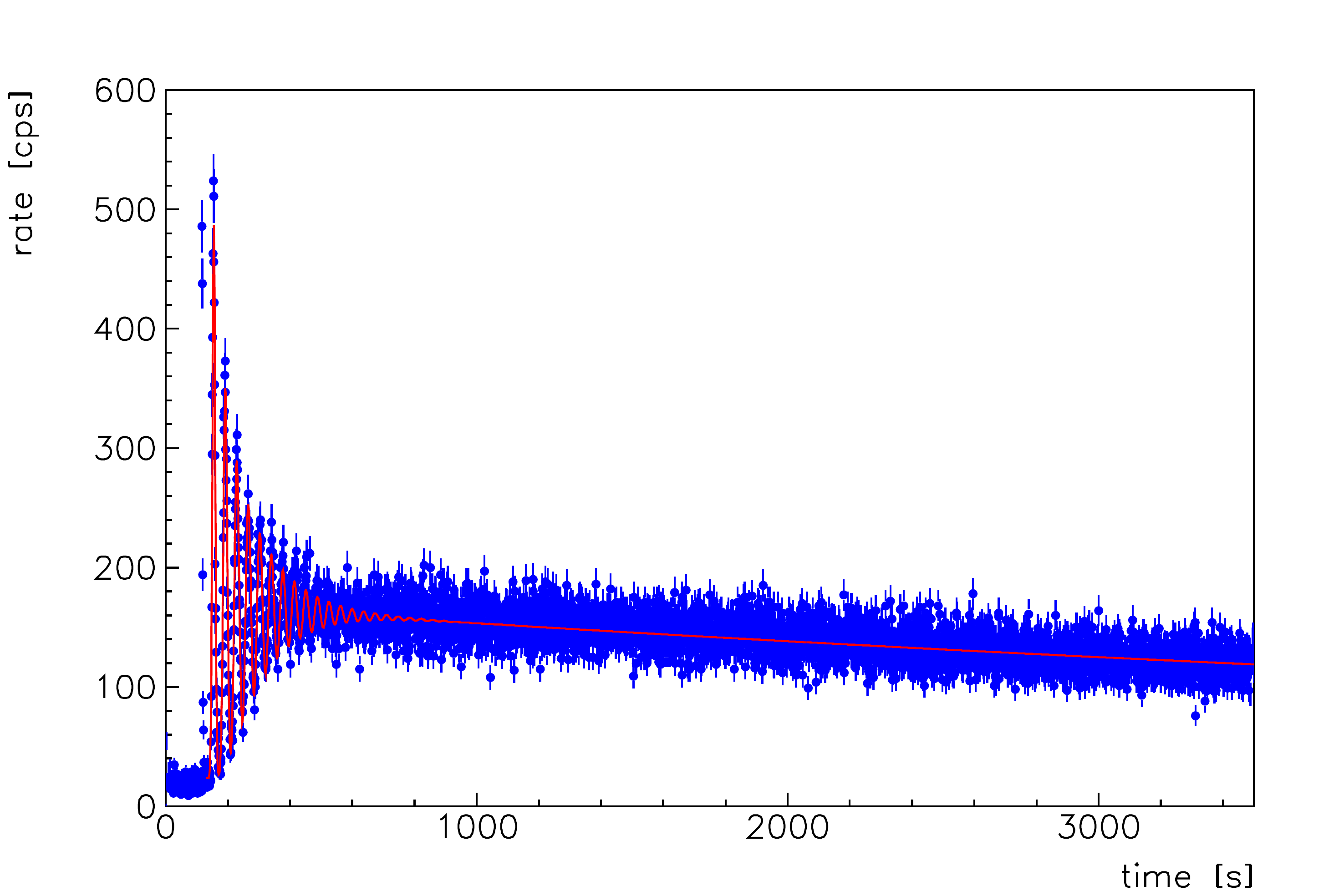}
\caption{\textbf{\krm\ decay rate in closed loop after injecting \krm\ source at t$\approx118$\,s.} The experimental data points are shown in blue, while the described fit is shown in red. The broadening of the Gaussians for each passing of the \krm\ by the detector due to diffusion results finally in a purely exponential decrease of the count rate in the \krm -decay detector.}
    \label{fig:timeevolution2}
\end{figure}

From the fit, the period of the oscillation has been determined to be $T_{per}=36.9\,\mathrm{s}\pm0.1\,\mathrm{s}$, which is the time needed by gas particles to perform one circle. This value can be compared to the measurement of the gas circulation by the flow controller. The volume of the system has been estimated to be $\approx 3000$\,cm$^{3}$ (with an uncertainty of 6\%) by using a known volume of gas and measuring the pressure decrease after expanding it to the system. With a forced gas flow of 8\,slpm and a gas pressure of 1\,bar xenon, the time for one cycle can be calculated to 22.5\,s. This is a huge discrepancy, which allows to determine the quality of the flow controller. Since this device is specified for high flow rates up to 200\,slpm, its accuracy at low flow rates is limited. The company is claiming an accuracy of $\pm1\%$ of full scale \cite{MKS}, i.e. 2\,slpm. That gives a relative uncertainty of 25\% at 8\,slpm on the reading of the flow. Therefore, the $^{83\mathrm{m}}$Kr decay allows for 
a cross 
calibration of the device, assuming that the measurement of the gas volume is correct.

Since the valve to the $^{83}$Rb source was closed, the amount of $^{83\mathrm{m}}$Kr particles inside the xenon decreases with the half-life of $^{83\mathrm{m}}$Kr. From the fit of the function $f(t)$, the $^{83\mathrm{m}}$Kr half-life was calculated to be 1.56\,h $\pm$ 0.04\,h. In the figure \ref{fig:timeevolution2}, the complete evolution of the detector signal in time is shown.  The measured half-life is slightly lower compared to the published value of 1.83\,h. This can be explained by diffusion of $^{83\mathrm{m}}$Kr into these parts of the system, where no circulation takes place, and by \krm\ adsorption at the system walls. The former can happen for example at pipes that lead to other subsystems.

\begin{figure}[!!!h]
\center
\includegraphics[width=0.9\textwidth]{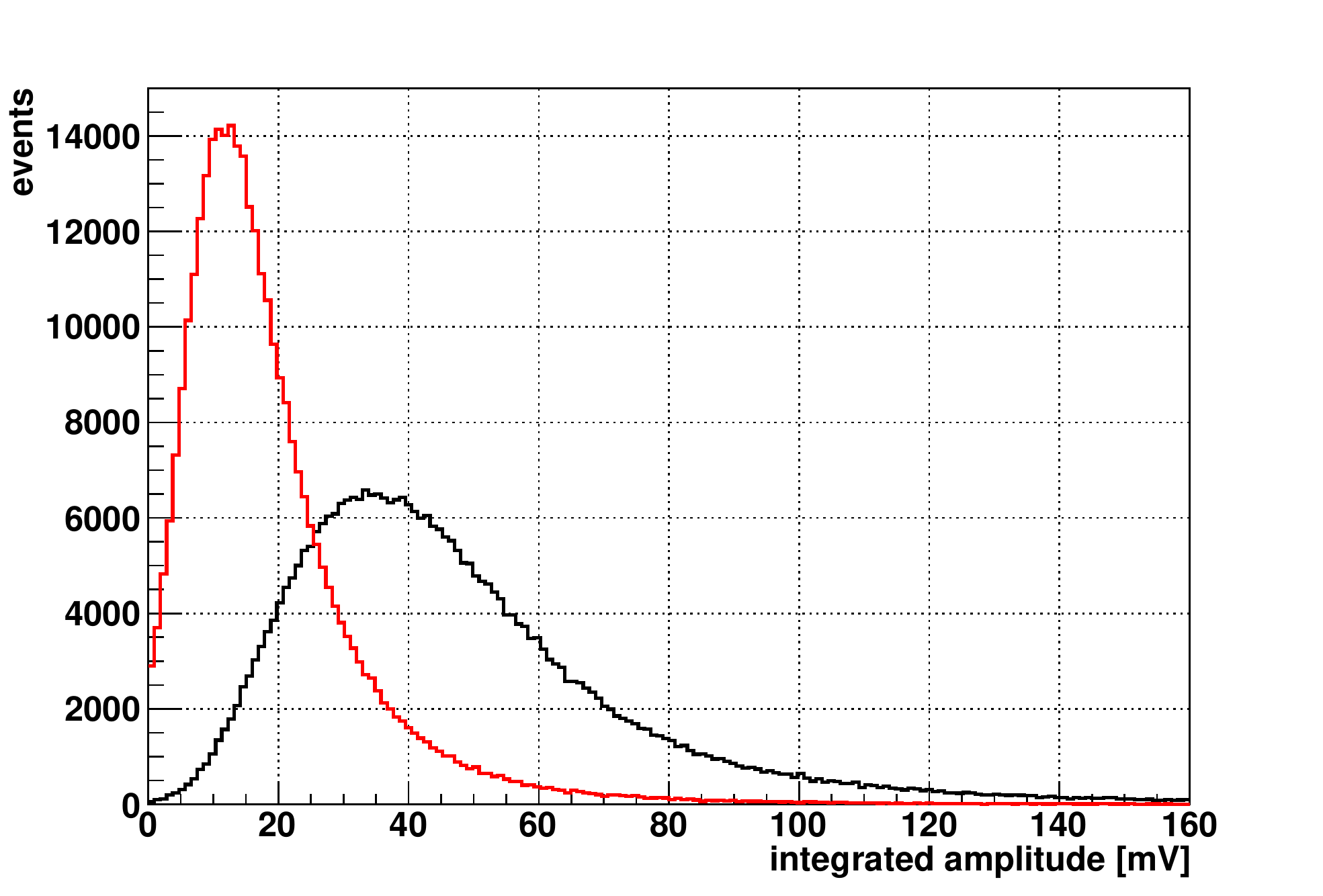}
\caption{\textbf{Integrated spectrum of the double coincidence signals.} The pulses of the first and second transition were selected from the measurement and integrated. The first transition at $\sim32$\,keV peaked at $\approx 35$\,mV (black), while the second transition at $\sim9$\,keV peaked at $\approx 12$\,mV (red).}
    \label{fig:integrSpectrum}
\end{figure}

Besides the time evolution of the rate, we could also determine energy spectra by integrating the signals and distinguishing between the two decay transitions of the $^{83\mathrm{m}}$Kr. After integrating the main signal, we identified and integrated the signal of the second transition as well. The result is shown in the figure \ref{fig:integrSpectrum}.

The black distribution is related to the first transition of 32\,keV while the red distribution is related to the second transition at 9\,keV.
As one can see, the energy resolution is limited: the peaks are very wide. For the 32\,keV transition the full width at half maximum is $FWHM\approx42$\,mV, while for the 9\,keV transition $FWHM\approx18$\,mV. Both distributions are overlapping each other. This can be explained by the high non-uniform light collection efficiency, which is limited by the usage of only one PMT together with a non-optimized PTFE reflector geometry. The ratio of the energies of the two transitions (see figure \ref{fig:rb_decay}) can be calculated to E$_{\gamma, 32keV}$/E$_{\gamma, 9keV}= 3.42$. From figure \ref{fig:integrSpectrum} the mean of the 32\,keV transition M$_{\gamma, 32keV}$ has been determined to M$_{\gamma, 32keV}$ = 47.19\,mV $\pm$ 0.05\,mV (the related median to the distribution is $\approx42$\,mV). For the 9\,keV transition the mean M$_{\gamma, 9keV}$ has been determined to be M$_{\gamma,9keV} =18.94$\,mV $\pm$ 0.03\,mV (the related median is $\approx 16$\,mV). The ratio of the mean (peak, median) values of the 
two measured transitions of 32~keV and 9~keV of figure 10 of $\approx 2.5$ ($\approx 2.9$, $\approx 2.6$) is always a bit lower than the expected value of 3.42. The origin of this discrepancy is very likely a non-linearity in the energy scale due to differences of the electron range and gamma mass attenuation length (see figure \ref{fig:attenuation}) and the imperfections of the light collection efficiency, a fact which is also reflected in the large width of the distributions.

\section{Application: testing of separation efficiencies in cryogenic distillation}

\begin{figure}[!!!h]
\center
\includegraphics[width=1.0\textwidth]{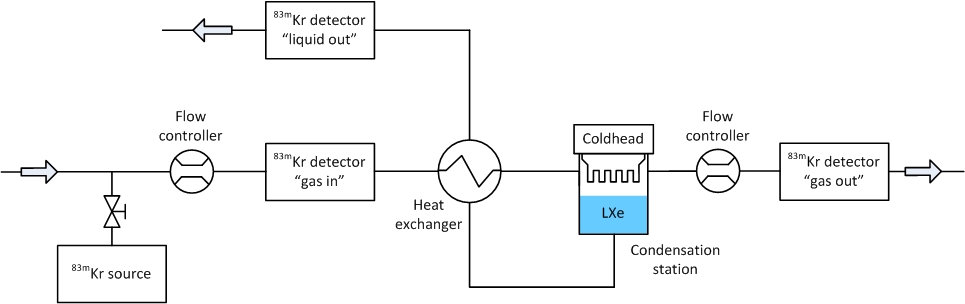}
\caption{\textbf{Scheme of the separation system.} The simplified scheme of the setup that was used to test the separation efficiency of one distillation stage.}
    \label{fig:preseparator}
\end{figure}

This tracer method was used for testing the separation efficiency of a cryogenic distillation stage. The scheme of the setup is shown in figure \ref{fig:preseparator}. 
The xenon is liquefied at temperature of $-98^{\circ}$C on a copper cooling block equipped with thin cooling fins to maximize the condensation surface. To keep the xenon clean, the condensation station is made from oxygen-free high thermal conductivity (OFHC) copper as well as stainless steel (alloy 316L). In addition, all parts have been cleaned in an ultra-sonic bath.  The cooling was performed with a cryo cooler (type CP50, Oerlikon-Leybold) operated at its maximum cooling power of 100\,W, while the temperature of $-98^{\circ}$C was stabilized with a heater cartridge of variable heating power controlled by a temperature controller (Model 336, Lakeshore).
The condensation station has two outputs, one for collecting xenon from the liquid phase and the other, located above the liquid level, for extraction xenon vapors. The incoming flow is controlled with a mass flow controller (type 1479B, MKS) with a maximum flow rate $q=20$\,slpm with an error of $\delta q=0.2$\,slpm, which is 1\% of the maximal flow rate.
The flow at the gas out line is controlled using a second flow controller of the same type (1479B) and the gas is fed back to the circuit. Since the circulation is driven in a closed loop, the flow at the liquid out line is defined by the inlet flow and the gas out flow in equilibrium.
This construction allows to control the fraction of xenon passing the gas out line relative to the liquid out line. Increasing the gas out flow would lead to a decrease in the liquid out flow due to particle conservation in the closed loop.

\begin{figure}[!!!h]
\center
\includegraphics[width=0.85\textwidth]{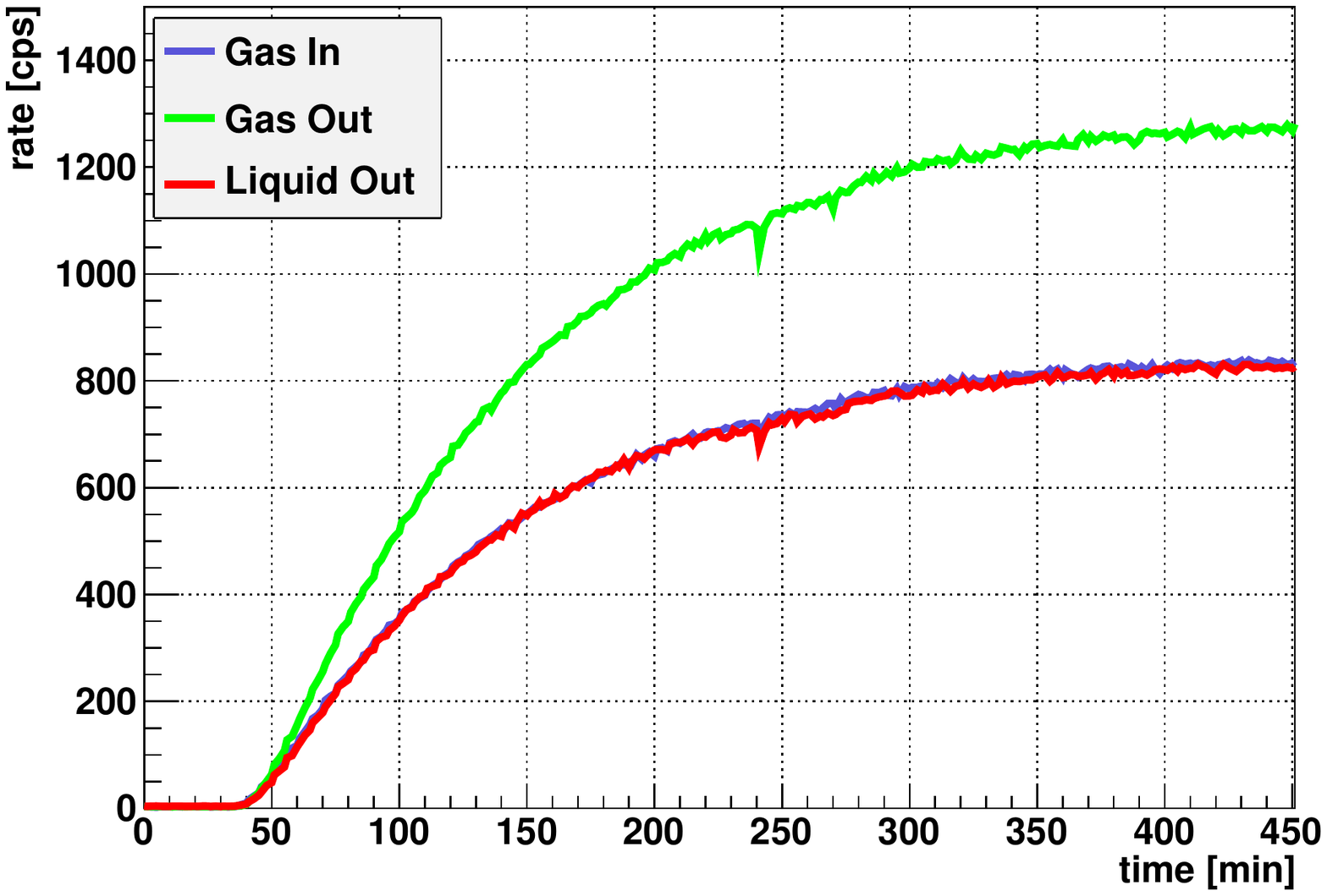}
\caption{\textbf{Rates measured by the $^{83\mathrm{m}}$Kr decay detectors with the system filled and circulated with gaseous xenon only.} The plot shows the rate evolution on the three $^{83\mathrm{m}}$Kr decay detectors for the system without any liquid load inside the condensation station. The valve to the \rb\ generator is opened at $\approx 2200$\,s followed by equal distribution of $^{83\mathrm{m}}$Kr in the whole system. The blue and the red lines lay closely together.}
    \label{fig:rates_in_GXe}
\end{figure}

There are three $^{83\mathrm{m}}$Kr decay detectors mounted inside the system, one at the inlet and by one at each outlet. The $^{83\mathrm{m}}$Kr particle flow can be monitored for different flow configurations with these detectors in combination with the flow information and pressure readings at different points of the system.

\begin{figure}[!!!h]
\center
\includegraphics[width=1.1\textwidth]{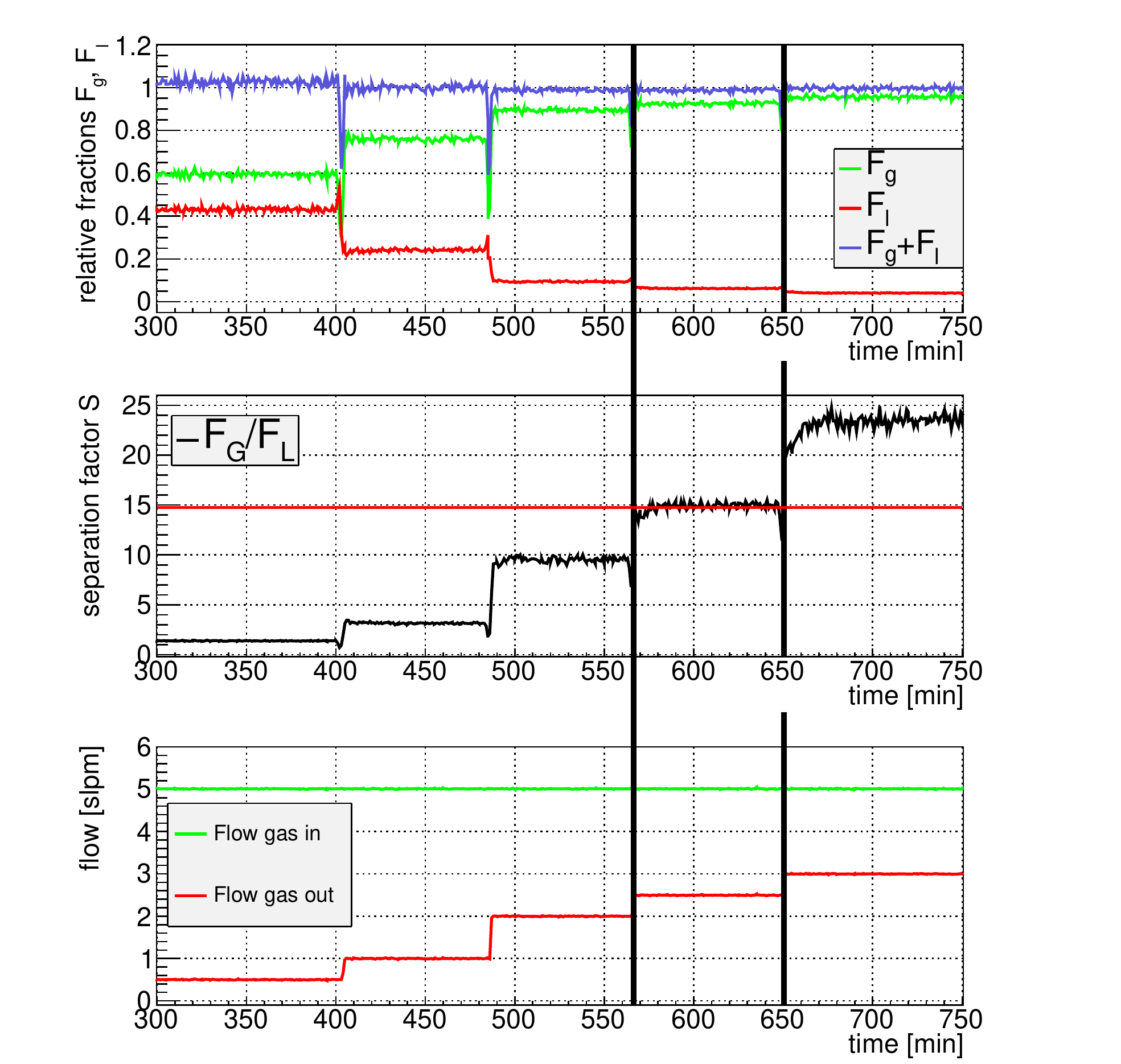}
\caption{\textbf{Krypton from xenon separation in the condensation station filled with dual phase xenon as measured by the \krm\ tracer method with the \krm -decay detectors.} Top: relative \krm\ flow rates according to the definition of $F_{g}$ and $F_{l}$, middle : separation factor S, bottom: flow rates. The steps in the rates are due to different flow ratios between gaseous and liquid outputs.}
    \label{fig:Presep_130124_pap}
\end{figure}

In figure \ref{fig:rates_in_GXe} the rates on all three $^{83\mathrm{m}}$Kr decay detectors were measured at room temperature (T$=20^{\circ}$C). The temperature at the cryo cooler is far above the xenon liquefaction temperature and no liquid level affects the gas flow. The $^{83\mathrm{m}}$Kr is distributed in the whole system, reaching an equilibrium. These measurements are necessary to determine the relative detector efficiencies. As one can see, the detector at the gas out line displays much higher rate than the other two detectors, although $^{83\mathrm{m}}$Kr is equally distributed in the system. These differences in detector efficiency must be taken into account in the further description (see equation \ref{eq:flow}).

The reservoir was then filled with certain amount of liquid xenon, by turning on the cryo cooler. For the measurement shown in figure \ref{fig:Presep_130124_pap}, the inventory was $\approx 360$\,ml of liquid xenon. Further measurements at different liquid volumes are presented in \cite{rosendahl1}, together with a more detailed description of the setup. During the performance tests, the temperature at the condensing station was stabilized to $-98^{\circ}$C and the circulating xenon gets liquefied. 

In order to determine the separation efficiency, it is necessary to calculate the $^{83\mathrm{m}}$Kr particle balance in the system from the measured rates at the detectors.

In general, the $^{83\mathrm{m}}$Kr particle flux $(\dot{N}(^{83\mathrm{m}}\mathrm{Kr}))$ can be expressed by the product of concentration c$(^{83\mathrm{m}}\mathrm{Kr})$ and the total particle flux $\dot{N}_{Tot}$. While the concentration is proportional to the rate $r$ at the detector, the total particle flux $q$ is measured by the flow controller. 

\begin{eqnarray}
 \dot{N}(^{83\mathrm{m}}\mathrm{Kr})=c(^{83\mathrm{m}}\mathrm{Kr}) \cdot \dot{N}_{Tot} = \frac{r\cdot q}{p}\cdot \frac{\tau}{V_{det}}
 \label{eq:flow}
\end{eqnarray}

where $\tau=T_{1/2}/\ln(2)$ is the mean life time of $^{83\mathrm{m}}$Kr and  $V_{det}$ is the effective detector volume that reflects the individual detector efficiencies. The pressure $p$ is measured by capacitance manometer (PTU-S, Swagelok)

In the next step, one can define the ratio between the measured signal of outgoing  $^{83\mathrm{m}}$Kr atoms at the gas outlet and ingoing  $^{83\mathrm{m}}$Kr atoms at the gas inlet $F_{g}$:

\begin{eqnarray}
  F_{g}=\frac{\dot{N}(^{83\mathrm{m}}\mathrm{Kr})_{gout}}{\dot{N}(^{83\mathrm{m}}\mathrm{Kr})_{gin}}=\frac{(c(^{83\mathrm{m}}\mathrm{Kr}) \cdot \dot{N}_{Tot})_{gout}}{(c(^{83\mathrm{m}}\mathrm{Kr})\cdot\dot{N}_{Tot})_{gin}}=\frac{r_{gout}\cdot q_{gout}\cdot p_{gin}}{r_{gin}\cdot q_{gin}\cdot p_{gout}}\cdot \frac{V_{det, gin}}{V_{det, gout}} 
\label{eq:fg}
\end{eqnarray}

The same is valid for the ratio between the measured signal of outgoing $^{83\mathrm{m}}$Kr atoms at the liquid out to the ingoing $^{83\mathrm{m}}$Kr atoms $F_{l}$ except for a correction factor $T_{Res}$.

\begin{eqnarray}
 F_{l}=\frac{\dot{N}(^{83\mathrm{m}}\mathrm{Kr})_{lout}}{\dot{N}(^{83\mathrm{m}}\mathrm{Kr})_{gin}}=\frac{(c(^{83\mathrm{m}}\mathrm{Kr}) \cdot \dot{N}_{Tot})_{lout}}{(c(^{83\mathrm{m}}\mathrm{Kr}) \cdot \dot{N}_{Tot})_{gin}}=\frac{r_{lout}\cdot q_{lout}\cdot p_{gin}}{r_{gin}\cdot q_{gin}\cdot p_{lout}}\cdot\frac{V_{det, gin}}{V_{det, lout}} \cdot \,T_{Res}
\label{eq:fl}
\end{eqnarray}

The factor $T_{Res}$ takes the half-life of $^{83\mathrm{m}}$Kr into account and is very important for long residence times of $^{83\mathrm{m}}$Kr atoms in the liquid phase. The separation factor can be then written as: $S=F_{g}/F_{l}$. 
 A detailed description of the derivation of equations \ref{eq:fg} and \ref{eq:fl} as well as the determination of $T_{Res}$ will also be given in \cite{rosendahl1}.

In figure \ref{fig:Presep_130124_pap} the time evolution of $F_{g}$ and $F_{l}$ is shown in the top plot, while the corresponding separation factor $S$ is shown in the middle one. During one measurement, different values for the gas out flow have been applied, as illustrated in the bottom plot. The black marked box shows the period, where an equal flow rate of 2.5\,slpm at both outputs has been set. Since krypton is the more volatile component compared to xenon (at 2\,bar the condensation temperature is $-143.98^{\circ}$C), it should preferably stay in the gas phase, what could be verified. At equal flow rates, the separation factor has been determined to be S $\approx15$ for the given conditions.

\section{Conclusion}

The suitability of $^{83\mathrm{m}}$Kr as a radioactive tracer in xenon gas has been demonstrated for the characterization of a gas purification system and a single stage distillation facility. For this application, custom-made detectors based on PMTs were designed and tested. They allow to measure the xenon scintillation light that is produced from the interactions of X-rays, $\gamma$\textquoteright s and electrons of the $^{83\mathrm{m}}$Kr decay with gaseous xenon. This method allows the characterization of gas circulation systems as described in chapter \ref{sec:gas_system}. A circulation velocity of $T_{per}=37\,\mathrm{s}$ has been measured and allows for a cross calibration of commercial flow controllers. In a second application, the separation efficiency of a single stage distillation system was demonstrated on one measurement that revealed separation factor of  S $\approx15$. As it had been expected, krypton as the more volatile component preferred to stay in the gas phase.
This technique can also be used to monitor the stability of larger distillation systems. Furthermore, the signal losses due to the radioactive decay can be used to determine the residence time of $^{83\mathrm{m}}$Kr atoms in such systems. Besides xenon, this method is applicable also to other noble gases, but that would require development of special detectors to monitor the scintillation light of different wavelengths.

\acknowledgments 

The R\&D for this tracer method are funded by DFG (WE 1843/7-1) and by Helmholtz Alliance of Astroparticle Physics HAP. Different aspects of the project and the contributions to the XENON1T experiment are supported by DFG Gro\ss ger\"ate (INST 211/528-1 FUGG, funded together with the state NRW and University of M\"unster) and by BMBF (05A11 PM1). The preparation of the $^{83\mathrm{m}}$Kr sources is supported by the Czech Science Foundation, GACR - grant number P203/12/1896 and by the CANAM project, funded by the Ministry of Education, Youth and Sport of the the Czech Republic (project no. LM2011019).

\end{document}